\documentclass[10pt,twocolumn,preprintnumbers,superscriptaddress,nofootinbib,aps,prd]{revtex4-2}

\usepackage{graphicx}
\usepackage{amsmath}
\usepackage{srcltx}
\usepackage[utf8]{inputenc}
\usepackage[colorlinks]{hyperref}
\usepackage{url}
\usepackage{times}
\usepackage{amssymb}
\newcommand{\vev}[1]{\langle {#1} \rangle}
\newcommand{\lsim}{\lesssim}
\newcommand{\gsim}{\gtrsim}

\newcommand{\eq}[1]{Eq.~(\ref{#1})}

\newcommand{\ord}[1]{\mathcal{O}{(#1)}}
\newcommand{\beq}{\begin{equation}}
\newcommand{\eeq}{\end{equation}}
\newcommand{\bea}{\begin{eqnarray}}
\newcommand{\eea}{\end{eqnarray}}

\newcommand{\etaw}{\eta_{\rm w}}
\newcommand{\mw}{m_{\eta \rm w}}
\newcommand{\fw}{f_{\eta \rm w}}

\newcommand{\mP}{M_{\rm P}}

\newcommand{\appropto}{\mathrel{\vcenter{
  \offinterlineskip\halign{\hfil$##$\cr
    \propto\cr\noalign{\kern2pt}\sim\cr\noalign{\kern-2pt}}}}}

\begin{document}

\pagestyle{plain}

\title{\boldmath  Astrophysical Consequences of an Electroweak $\eta_{\rm w}$ Pseudo-Scalar}

\author{Hooman Davoudiasl}
\email{hooman@bnl.gov} 
\affiliation{High Energy Theory Group, Physics Department \\ Brookhaven National Laboratory,
Upton, NY 11973, USA}


\begin{abstract}
	
Recently, it has been suggested that the spectrum of physical states in the Standard Model may include an ultralight pseudo-scalar, denoted by $\eta_{\rm w}$, in analogy with the $\eta'$ state arising from the strong interactions.  We find that typical expectations for the properties of $\eta_{\rm w}$ get challenged by astrophysical constraints on the couplings of ultralight bosons. Our strongest limit sets a lower bound of $\ord{\rm 100~TeV}$ on the decay constant of the hypothesized pseudo-scalar.  We also briefly discuss whether $\eta_{\rm w}$ could be a dark matter candidate, or the origin of dark energy, but conclude that those identifications appear unlikely.  Given the important implications of a potentially overlooked $\eta_{\rm w}$ state for a more complete understanding of the electroweak interactions and a fundamental description of Nature, further theoretical and phenomenological investigations of this possibility and its associated physics are warranted.        
\end{abstract}
\maketitle


It has recently been proposed that the dynamics of the electroweak sector leads to the emergence of an ultralight pseudo-scalar, denoted by $\etaw$, whose mass arises from the anomaly of the global $U(1)_{\rm B+L}$ symmetry of the $B+L$ charge, where $B$ and $L$ are the baryon and lepton numbers, respectively \cite{Dvali:2024zpc,Dvali:2025pcx}.  The gist of this proposal is that since the vacuum angle $\theta_{\rm w}$ associated with the $SU(2)$ gauge interactions of the Standard Model (SM) is not physical \cite{Anselm:1992yz,Anselm:1993uj,FileviezPerez:2014xju}, there must exist a pseudo-scalar, the aforemnetioned $\etaw$, that is responsible for its removal.  This is akin to the same effect in the strong interactions governed by the SM quantum chromodynamics (QCD), in the presence of a massless quark that would make the analogous angle $\theta_{\rm QCD}$ unphysical.  Within QCD, this effect would be ascribed to the infrared contribution of the observed pseudo-scalar $\eta'$ \cite{Witten:1979vv}, whose relatively large mass $\sim 1$~GeV originates from the anomaly of the axial $U(1)_A$ symmetry. 

It is fair to say that the status of $\etaw$ as a firm prediction of the SM is not well-established and some of its fundamental properties, like mass and coupling, remain to be confidently predicted  \cite{Dvali:2025pcx}.  It has also been suggested that perhaps this state is not ultralight and is in fact a superposition of hydrogen and anti-hydrogen \cite{Cacciapaglia:2025xmr}.  This is an interesting idea which may play a role in the phenomenology of the relevant physics.  However, as pointed out in Ref.~\cite{Dvali:2025pcx}, it does not seem to fit the assumed features of a particle associated with the removal of $\theta_{\rm w}$ from the electroweak interactions.  This association would make sense if one could treat $\etaw$ as a phase, {\it i.e.} a Goldstone mode, and it is not obvious that a composite atomic state can fulfill this role. 

We find that the theoretical arguments put forth in Refs.~\cite{Dvali:2024zpc,Dvali:2025pcx} -- in order to support the presence of $\etaw$ within the SM -- provide good motivation for examining this intriguing hypothesis further.  The emergence of $\etaw$ from the SM electroweak interactions would have significant import for our understanding of fundamental physics.  In light of this circumstance, it is interesting to examine the feasibility of this hypothesis from various angles.  Here, we will simply  assume that the $\etaw$ pseudo-scalar described in Refs.~\cite{Dvali:2024zpc,Dvali:2025pcx} is a prediction of the SM.  We will then examine to what extent current astrophysical bounds, in particular those related to stellar emission of new light particles, can constrain the basic properties of $\etaw$.

Let us denote the mass of $\etaw$ by $\mw$ and its decay constant by $\fw$.  What do we expect for the value of $\mw$?  This mass scale is assumed to be generated by electroweak instantons due to the $B+L$ anomaly in the SM.  Therefore, it seems reasonable to take the Higgs vacuum expectation value $v=246$~GeV as the typical mass scale $M$ of the relevant dynamics.  However, the associated instanton effects are suppressed by $e^{-2\pi/\alpha_2(v)} \sim 10^{-80}$, where $\alpha_2 \equiv g_2^2/(4\pi)$ and $g_2 \approx 0.65$ is the SM $SU(2)$ coupling at the weak scale $\sim v$.  On general grounds, we may then expect 
\beq
\mw^2 \lsim e^{\frac{-2\pi}{\alpha_2(M)}} M^2  
\label{mw2}
\eeq
which for $M\sim v$ yields $\mw \lsim 10^{-29}$~eV.  This is a very tiny mass, making $\etaw$ a genuine ultralight scalar.  Note that even for an unlikely value $M\sim \mP\sim 10^{19}$~GeV, {\it i.e.} near the Planck scale, we get $\mw\sim 10^{-39}$~eV, which is even deeper in the ultralight regime, due to the falling value of $\alpha_2$ with energy in the SM.  Here, we have used $\alpha_2 (\mP) \approx 1/50$, given by the 1-loop running formula 
\beq
\frac{1}{\alpha_2(\mu)} = \frac{1}{\alpha_2(\mu_0)} + \frac{\beta_0}{2 \pi} \ln(\mu/\mu_0)\,,
\label{running}
\eeq
where $\beta_0 = 19/6$ in the SM (see, Ref.~\cite{McLerran:2012mm}, for example).  We thus conclude that, under any reasonable assumptions consistent with the assumed origin of the $\etaw$ state, $\mw$  is extremely small and within the kinematic reach of stellar physical processes.

Next, we will focus on the decay constant $\fw$.  In Ref.~\cite{Dvali:2025pcx}, no definitive value is derived, but $\fw\lsim v$ is considered as reasonable, in light of theoretic considerations.  In the scenario where the physics of $\etaw$ is entirely contained in the SM, this scalar is associated with a condensate that emerges from the dynamics of the 't Hooft vertex through the electroweak $SU(2)$ instantons \cite{Dvali:2024zpc,Dvali:2025pcx}.  This condensate is made up of quark $Q$ and lepton $\ell$ doublets
\beq
\vev{(QQQ\ell)^{N_f}} \propto e^{\frac{-2\pi}{\alpha_2(v)}} v^{6 N_f},
\label{condensate}
\eeq 
where $N_f=3$ is the number of generations in the SM \cite{Dvali:2024zpc,Dvali:2025pcx}.  The above condensate corresponds to the order parameter for the spontaneous breaking of the global $U(1)_{\rm B+L}$ symmetry and hence the decay constant $\fw$ depends on its size.  This is in analogy with the QCD chiral condensate $\vev{\bar q_L q_R}$ that sets the pion decay constant $f_\pi$.  Hence, given the instanton suppression in \eq{condensate}, it is reasonable to assume that $\fw \lsim v$ from electroweak non-perturbative dynamics, which we will adopt as the typical expectation, consistent with arguments put forth in Ref.~\cite{Dvali:2025pcx}.

No matter what the actual value of $\fw$ is, it sets the coupling of $\etaw$ to the SM $SU(2)$ gauge fields, through the anomaly, by
\beq
\frac{\alpha_2}{8\pi}\frac{\etaw}{\fw} W^a_{\mu\nu}\tilde{W}^{a\mu\nu}\,,
\label{etawWW}
\eeq      
where $W^a_{\mu\nu}$ is the field strength tensor, $\tilde{W}^a_{\mu\nu}$ is its dual, and $a=1,2,3$ is the adjoint index of $SU(2)$.  Upon electroweak symmetry breaking, the above yields an interaction between $\etaw$ and the weak gauge bosons, $W^\pm$ and $Z$.  A similar coupling to photons is vanishing due to anomaly cancellation, and the direct coupling to fermions is exponentially suppressed by the electroweak instanton rate \cite{Dvali:2024zpc}.  

The interaction in \eq{etawWW} leads to a loop-induced coupling to quarks and leptons, which can potentially provide missing energy signals in meson decays \cite{Dvali:2024zpc}.  The $W^\pm$ interaction also generates a 1-loop coupling to photons, however this coupling is proportional to $\mw^2/m_W^2$, where $m_W=80.4$~GeV is the $W^\pm$ boson mass, and hence extremely suppressed; see Ref.~\cite{Bauer:2017ris} for an effective field theory derivation (a similar result has also been reported for the two-loop majoron coupling to photons in Ref.~\cite{Heeck:2019guh}).   Therefore, we will only consider the interactions of $\etaw$ with fermions, in relation to the astrophysical bounds below.  However, first we need to make an estimate of the coupling strengths.  Given the uncertainty regarding the precise form of the physics that describes $\etaw$, it suffices to use estimates that are correct up to $\ord{1}$ factors, in what follows.  

\begin{figure}[t]\vskip0.25cm	
	\includegraphics[width=0.9\columnwidth]{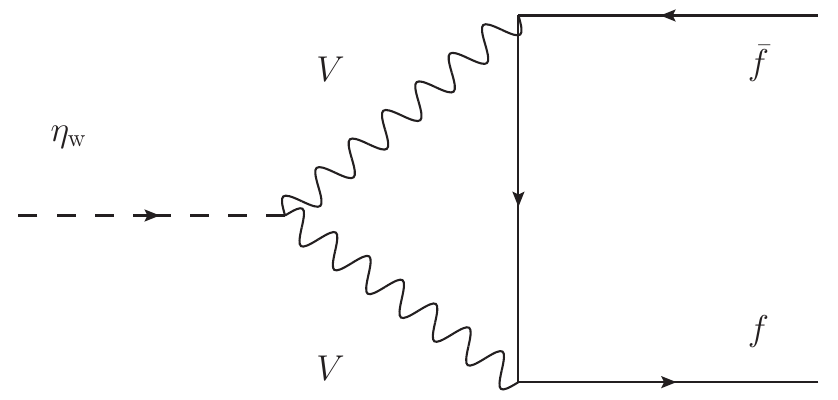}
	\caption{A representative 1-loop process that couples $\etaw$ to a fermion $f$.  Here, $V$ denotes a heavy $SU(2)$ gauge boson.}
	\label{W-loop}
\end{figure}

Let us consider the class of diagrams represented in Fig.\ref{W-loop}.  Here, $V$ is a heavy electroweak vector boson $W^\pm, Z$ and $f$ is a fermion of the SM.  We will use the result of Ref.~\cite{Bauer:2017ris}, where the contribution of such processes to the coupling between an  axion-like-particle and fermions was calculated. Let us parameterize the coupling to a charged lepton  $l$ by 
\beq
c_{ll} \frac{m_l}{\fw} \, \etaw\, \bar l \gamma_5\, l\,,
\label{etawll}
\eeq    
where $c_{ll}$ is a numerical coefficient and $m_l$ is the charged lepton mass.  From the results of Ref.~\cite{Bauer:2017ris}, we have 
\beq
c_{ll} = \left(\frac{3 \alpha_2}{8\pi}\right)^2 \ln\left(\frac{\Lambda^2}{m_W^2}\right)\,,
\label{cll}
\eeq
where $\Lambda$ is a cutoff scale which we set by $\Lambda\gsim m_W$, given that the loop momenta corresponding to the process in Fig.~\ref{W-loop} are governed by scales of order the $W$ mass.  We may require that 
\beq
\Lambda = 4\pi \fw\,, 
\label{Lambda}
\eeq
as a natural cutoff scale for the $\etaw$ dynamics.  In this case, we find 
$\fw\gsim 6$~GeV, in a regime that may be suggested by the general arguments in Ref.~\cite{Dvali:2025pcx}.

Using the above relations, we can estimate the coupling of $\etaw$ to electrons 
\beq
y_e \,\etaw\, \bar e \gamma_5 e\,.
\label{etawee}
\eeq   
Setting $\ln(\Lambda^2/m_W^2)\sim 1$ in \eq{cll}, we find 
\beq
y_e \sim (2\times 10^{-5})\frac{m_e}{\fw}\,,
\label{ye}
\eeq
where $m_e\approx 0.511$~MeV is the electron mass.  Given the ultralight nature of $\etaw$, stellar cooling bounds apply.  Here, the most stringent bound comes from globular cluster red giant energy loss, with $y_e\lsim 10^{-13}$ \cite{Marsh:2015xka,Raffelt:1994ry,Raffelt:2006cw,Caputo:2024oqc}, which implies 
\beq
\fw \gsim 100~\text{TeV}.\quad (\text{Red giants})  
\label{Redgiant}
\eeq
This bound pushes $\fw$ three orders of magnitude above the weak scale and seems to be in tension with an electroweak origin for $\etaw$ \cite{Dvali:2024zpc,Dvali:2025pcx}.

In recent years, the coupling of light bosons to muons have been considered in the context of astrophysics, in particular core collapse supernova explosions.  The investigations in Refs.~\cite{Bollig:2020xdr,Croon:2020lrf} suggest that the coupling of muons to a light pseudo-scalar is constrained by the Supernova 1987A to be $y_\mu \lsim \text{few}\times 10^{-9}$.  Following the  treatment applied to the electron coupling before, this would imply $\fw\gsim 1$~TeV, which is not competitive with the above red giant bound.  In general, the derivation of preceding bounds could be affected by the trapping of axions in the stellar environment.  However, that regime of parameters is generally ruled out by other considerations~\cite{Raffelt:2006cw}.  

In the case of first generation quarks, $f=u,d$, the diagram in Fig.~\ref{W-loop} provides a coupling between $\etaw$ and nucleons $N$ 
\beq
y_N \,\etaw\, \bar N \gamma_5 N\,,
\label{etawNN}
\eeq   
of strength $y_N$.  Along the lines of the preceding discussion, we may roughly estimate $y_N$ by 
\beq
y_N\sim \left(\frac{3 \alpha_2}{8\pi}\right)^2\frac{m_{q}}{\fw}\,,
\label{yN}
\eeq
where $m_q \sim 5$~MeV is a typical valence quark mass inside a nucleon.  Note that the above estimate  is at the level of quarks, but we expect that up to $\ord{1}$ effects this can be taken as the nucleon coupling.  For nucleon couplings, we have $y_N\lsim 10^{-9}$ \cite{Marsh:2015xka,Raffelt:2006cw,Caputo:2024oqc}.  Using the above estimate for $y_N$, we thus find that $\fw\gsim 100$~GeV would be sufficient, which may fit within a weak scale scenario for $\etaw$, but is quite a bit less stringent than the one in \eq{Redgiant}.

Here, we note that the non-perturbative nature of the dynamics governing the properties of $\etaw$ could in principle defy naive expectations for the size of $\fw$.  Nonetheless, the above astrophysical constraints seem to suggest that if $\etaw$ is present in the SM spectrum of states, its underlying dynamics may be more involved than what may be inferred from straightforward arguments.

Let us now turn to the question of dark matter (DM), which comprises about 25\% of cosmic energy density \cite{ParticleDataGroup:2024cfk}.   The identity of DM remains a mystery of great interest to both particle physics and cosmology.  A great deal of effort, in theory as well as experiment, is devoted to finding the answer to this question, which is one of the main motivations in searching for new physics beyond the SM.  This is because no particle, elementary or otherwise, within the SM domain is believed to account for the DM content of the Universe.  While neutrinos are weakly interacting and long-lived, they do not have the requisite mass and cosmological properties to establish the observed DM abundance and characteristics.  It has also been suggested that there may be a deeply bound long-lived state of six quarks \cite{Jaffe:1976yi} -- often called the  ``dibaryon" -- that could be DM \cite{Farrar:2003qy,Farrar:2017ysn}.  However, no such state has been observed and it is quite doubtful that it could be the dominant form of matter in the Universe \cite{Kolb:2018bxv,McDermott:2018ofd}.  

In light of the above considerations, it is important to ask whether $\etaw$ -- assuming that it arises without the need for physics outside of the SM -- can be a viable DM candidate.  As discussed before, a weak scale origin for this particle would suggest $\mw \sim 10^{-29}$~eV which is too light.  To see this, note that its de Broglie wavelength would be given by 
\beq
\lambda_{\rm w} \sim (\mw v_g)^{-1}\,,
\label{lw}
\eeq
where $v_g \lsim 10^{-3}$ is a typical galactic virial velocity.  The above equation then suggests $\lambda_{\rm w} \gsim 10^{32}~\text{eV}^{-1}\gsim$~Gpc, which is several orders of magnitude large than the size of a galaxy.  Thus, $\etaw$ cannot be confined within galactic structures, as observational data require.  Hence, it seems unlikely that $\etaw$ could be a viable DM candidate.   

A very light scalar could in principle provide an explanation for the observed accelerated expansion of the Universe~\cite{ParticleDataGroup:2024cfk} (see also Ref.~\cite{McLerran:2012mm} for an alternative view of the electroweak instanton contribution to the cosmological constant).  In this case, the field should be essentially static and not have any significant kinetic energy at the present epoch.  This amounts to requiring that $\mw \lsim H_0$, where $H_0\sim 10^{-33}$~eV is the present Hubble constant.  Then, the potential energy of $\etaw$ could provide the requisite ``dark energy" accelerating the expansion of the Universe, which we may estimate by $\rho_{\rm w} \sim \mw^2 \fw^2$.  However, if we take $\fw \sim 100$~TeV, as close as possible to the weak scale allowed by astrophysical bounds above, we would get $\rho_{\rm w} \lsim 10^{-34}$~eV$^4$.  The dark energy stored in the $\etaw$ field is thus expected to be exponentially small compared to the value $\sim 10^{-11}$~eV$^4$~\cite{ParticleDataGroup:2024cfk} inferred from observational data.  Getting close to the observed value requires $\fw\sim \mP$ for $\mw\sim H_0$.               
    
{\it Summary and concluding remarks:}---In this work, we have considered the astrophysical implications of a possible ultralight pseudo-scalar, the $\etaw$.  It has been argued in Refs.\cite{Dvali:2024zpc,Dvali:2025pcx} that this boson can emerge in the SM, without the need for any new ingredients, due to non-perturbative electroweak interactions and the $B+L$ anomaly.  The arguments that have led to this conclusion are subtle, but have close kinship with those that relate the observed $\eta'$ meson to QCD instantons and the axial anomaly.  As such, the case for $\etaw$ can be considered motivated and worth further scrutiny, as it can have profound implications for the physical spectrum of the SM particles and electroweak dynamics.

Our results indicate that typical assumptions about the basic physical properties of $\etaw$ are in tension with astrophysical constraints on ultralight bosons.  We found that avoiding excessive cooling of red giants results in the strongest such constraint, providing a lower bound $\fw \gsim 100$~TeV on the decay constant of a hypothetical $\etaw$.  This is roughly three or more orders of magnitude above the expectation $\fw \lsim 100$~GeV, assuming an electroweak origin.  

While the underlying non-perturbative dynamics of such a particle is non-trivial, the bounds on its coupling, as obtained in this work, challenge typical arguments for the emergence of $\etaw$ and its general properties.  Nonetheless, we view the possibility of a hitherto-ignored ultralight state of the SM sufficiently intriguing that further investigations of this topic seem well-justified.  If theoretical evidence for $\etaw$ gets more robust and difficult to refute, its detection would become an urgent experimental priority in fundamental physics.  In that case, our results suggest that astrophysical observations could provide a promising avenue for discovery of $\etaw$.


~
\begin{acknowledgments}
We thank M. Berbig for pointing out an error in an earlier version of our manuscript, regarding the estimation of the $\etaw$ coupling to photons.  This work is supported by the US Department of Energy under Grant Contract DE-SC0012704. 
\end{acknowledgments}

\bibliography{etaW.bib}

\begin{thebibliography}{22}%
\makeatletter
\providecommand \@ifxundefined [1]{%
 \@ifx{#1\undefined}
}%
\providecommand \@ifnum [1]{%
 \ifnum #1\expandafter \@firstoftwo
 \else \expandafter \@secondoftwo
 \fi
}%
\providecommand \@ifx [1]{%
 \ifx #1\expandafter \@firstoftwo
 \else \expandafter \@secondoftwo
 \fi
}%
\providecommand \natexlab [1]{#1}%
\providecommand \enquote  [1]{``#1''}%
\providecommand \bibnamefont  [1]{#1}%
\providecommand \bibfnamefont [1]{#1}%
\providecommand \citenamefont [1]{#1}%
\providecommand \href@noop [0]{\@secondoftwo}%
\providecommand \href [0]{\begingroup \@sanitize@url \@href}%
\providecommand \@href[1]{\@@startlink{#1}\@@href}%
\providecommand \@@href[1]{\endgroup#1\@@endlink}%
\providecommand \@sanitize@url [0]{\catcode `\\12\catcode `\$12\catcode
  `\&12\catcode `\#12\catcode `\^12\catcode `\_12\catcode `\%12\relax}%
\providecommand \@@startlink[1]{}%
\providecommand \@@endlink[0]{}%
\providecommand \url  [0]{\begingroup\@sanitize@url \@url }%
\providecommand \@url [1]{\endgroup\@href {#1}{\urlprefix }}%
\providecommand \urlprefix  [0]{URL }%
\providecommand \Eprint [0]{\href }%
\providecommand \doibase [0]{https://doi.org/}%
\providecommand \selectlanguage [0]{\@gobble}%
\providecommand \bibinfo  [0]{\@secondoftwo}%
\providecommand \bibfield  [0]{\@secondoftwo}%
\providecommand \translation [1]{[#1]}%
\providecommand \BibitemOpen [0]{}%
\providecommand \bibitemStop [0]{}%
\providecommand \bibitemNoStop [0]{.\EOS\space}%
\providecommand \EOS [0]{\spacefactor3000\relax}%
\providecommand \BibitemShut  [1]{\csname bibitem#1\endcsname}%
\let\auto@bib@innerbib\@empty
\bibitem [{\citenamefont {Dvali}\ \emph
  {et~al.}(2025{\natexlab{a}})\citenamefont {Dvali}, \citenamefont
  {Kobakhidze},\ and\ \citenamefont {Sakhelashvili}}]{Dvali:2024zpc}%
  \BibitemOpen
  \bibfield  {author} {\bibinfo {author} {\bibfnamefont {G.}~\bibnamefont
  {Dvali}}, \bibinfo {author} {\bibfnamefont {A.}~\bibnamefont {Kobakhidze}},\
  and\ \bibinfo {author} {\bibfnamefont {O.}~\bibnamefont {Sakhelashvili}},\
  }\bibfield  {title} {\bibinfo {title} {{Electroweak $\eta_{\rm w}$ meson}},\
  }\href {https://doi.org/10.1103/mxdm-wb1p} {\bibfield  {journal} {\bibinfo
  {journal} {Phys. Rev. D}\ }\textbf {\bibinfo {volume} {111}},\ \bibinfo
  {pages} {113002} (\bibinfo {year} {2025}{\natexlab{a}})},\ \Eprint
  {https://arxiv.org/abs/2408.07535} {arXiv:2408.07535 [hep-th]} \BibitemShut
  {NoStop}%
\bibitem [{\citenamefont {Dvali}\ \emph
  {et~al.}(2025{\natexlab{b}})\citenamefont {Dvali}, \citenamefont
  {Kobakhidze},\ and\ \citenamefont {Sakhelashvili}}]{Dvali:2025pcx}%
  \BibitemOpen
  \bibfield  {author} {\bibinfo {author} {\bibfnamefont {G.}~\bibnamefont
  {Dvali}}, \bibinfo {author} {\bibfnamefont {A.}~\bibnamefont {Kobakhidze}},\
  and\ \bibinfo {author} {\bibfnamefont {O.}~\bibnamefont {Sakhelashvili}},\
  }\bibfield  {title} {\bibinfo {title} {{$\eta_{\rm w}$-meson from topological
  properties of the electroweak vacuum}},\ }\href@noop {} {\  (\bibinfo {year}
  {2025}{\natexlab{b}})},\ \Eprint {https://arxiv.org/abs/2509.16043}
  {arXiv:2509.16043 [hep-th]} \BibitemShut {NoStop}%
\bibitem [{\citenamefont {Anselm}\ and\ \citenamefont
  {Johansen}(1993)}]{Anselm:1992yz}%
  \BibitemOpen
  \bibfield  {author} {\bibinfo {author} {\bibfnamefont {A.~A.}\ \bibnamefont
  {Anselm}}\ and\ \bibinfo {author} {\bibfnamefont {A.~A.}\ \bibnamefont
  {Johansen}},\ }\bibfield  {title} {\bibinfo {title} {{Baryon nonconservation
  in standard model and Yukawa interaction}},\ }\href
  {https://doi.org/10.1016/0550-3213(93)90060-3} {\bibfield  {journal}
  {\bibinfo  {journal} {Nucl. Phys. B}\ }\textbf {\bibinfo {volume} {407}},\
  \bibinfo {pages} {313} (\bibinfo {year} {1993})}\BibitemShut {NoStop}%
\bibitem [{\citenamefont {Anselm}\ and\ \citenamefont
  {Johansen}(1994)}]{Anselm:1993uj}%
  \BibitemOpen
  \bibfield  {author} {\bibinfo {author} {\bibfnamefont {A.~A.}\ \bibnamefont
  {Anselm}}\ and\ \bibinfo {author} {\bibfnamefont {A.~A.}\ \bibnamefont
  {Johansen}},\ }\bibfield  {title} {\bibinfo {title} {{Can electroweak theta
  term be observable?}},\ }\href {https://doi.org/10.1016/0550-3213(94)90392-1}
  {\bibfield  {journal} {\bibinfo  {journal} {Nucl. Phys. B}\ }\textbf
  {\bibinfo {volume} {412}},\ \bibinfo {pages} {553} (\bibinfo {year}
  {1994})},\ \Eprint {https://arxiv.org/abs/hep-ph/9305271}
  {arXiv:hep-ph/9305271} \BibitemShut {NoStop}%
\bibitem [{\citenamefont {Fileviez~Perez}\ and\ \citenamefont
  {Patel}(2014)}]{FileviezPerez:2014xju}%
  \BibitemOpen
  \bibfield  {author} {\bibinfo {author} {\bibfnamefont {P.}~\bibnamefont
  {Fileviez~Perez}}\ and\ \bibinfo {author} {\bibfnamefont {H.~H.}\
  \bibnamefont {Patel}},\ }\bibfield  {title} {\bibinfo {title} {{The
  electroweak vacuum angle}},\ }\href
  {https://doi.org/10.1016/j.physletb.2014.03.064} {\bibfield  {journal}
  {\bibinfo  {journal} {Phys. Lett. B}\ }\textbf {\bibinfo {volume} {732}},\
  \bibinfo {pages} {241} (\bibinfo {year} {2014})},\ \Eprint
  {https://arxiv.org/abs/1402.6340} {arXiv:1402.6340 [hep-ph]} \BibitemShut
  {NoStop}%
\bibitem [{\citenamefont {Witten}(1979)}]{Witten:1979vv}%
  \BibitemOpen
  \bibfield  {author} {\bibinfo {author} {\bibfnamefont {E.}~\bibnamefont
  {Witten}},\ }\bibfield  {title} {\bibinfo {title} {{Current Algebra Theorems
  for the U(1) Goldstone Boson}},\ }\href
  {https://doi.org/10.1016/0550-3213(79)90031-2} {\bibfield  {journal}
  {\bibinfo  {journal} {Nucl. Phys. B}\ }\textbf {\bibinfo {volume} {156}},\
  \bibinfo {pages} {269} (\bibinfo {year} {1979})}\BibitemShut {NoStop}%
\bibitem [{\citenamefont {Cacciapaglia}\ \emph {et~al.}(2025)\citenamefont
  {Cacciapaglia}, \citenamefont {Sannino},\ and\ \citenamefont
  {Turner}}]{Cacciapaglia:2025xmr}%
  \BibitemOpen
  \bibfield  {author} {\bibinfo {author} {\bibfnamefont {G.}~\bibnamefont
  {Cacciapaglia}}, \bibinfo {author} {\bibfnamefont {F.}~\bibnamefont
  {Sannino}},\ and\ \bibinfo {author} {\bibfnamefont {J.}~\bibnamefont
  {Turner}},\ }\bibfield  {title} {\bibinfo {title} {{Hiding in Plain Sight,
  the electroweak $\eta_W$}},\ }\href@noop {} {\  (\bibinfo {year} {2025})},\
  \Eprint {https://arxiv.org/abs/2509.15912} {arXiv:2509.15912 [hep-ph]}
  \BibitemShut {NoStop}%
\bibitem [{\citenamefont {McLerran}\ \emph {et~al.}(2012)\citenamefont
  {McLerran}, \citenamefont {Pisarski},\ and\ \citenamefont
  {Skokov}}]{McLerran:2012mm}%
  \BibitemOpen
  \bibfield  {author} {\bibinfo {author} {\bibfnamefont {L.}~\bibnamefont
  {McLerran}}, \bibinfo {author} {\bibfnamefont {R.}~\bibnamefont {Pisarski}},\
  and\ \bibinfo {author} {\bibfnamefont {V.}~\bibnamefont {Skokov}},\
  }\bibfield  {title} {\bibinfo {title} {{Electroweak Instantons, Axions, and
  the Cosmological Constant}},\ }\href
  {https://doi.org/10.1016/j.physletb.2012.05.057} {\bibfield  {journal}
  {\bibinfo  {journal} {Phys. Lett. B}\ }\textbf {\bibinfo {volume} {713}},\
  \bibinfo {pages} {301} (\bibinfo {year} {2012})},\ \Eprint
  {https://arxiv.org/abs/1204.2533} {arXiv:1204.2533 [hep-ph]} \BibitemShut
  {NoStop}%
\bibitem [{\citenamefont {Bauer}\ \emph {et~al.}(2017)\citenamefont {Bauer},
  \citenamefont {Neubert},\ and\ \citenamefont {Thamm}}]{Bauer:2017ris}%
  \BibitemOpen
  \bibfield  {author} {\bibinfo {author} {\bibfnamefont {M.}~\bibnamefont
  {Bauer}}, \bibinfo {author} {\bibfnamefont {M.}~\bibnamefont {Neubert}},\
  and\ \bibinfo {author} {\bibfnamefont {A.}~\bibnamefont {Thamm}},\ }\bibfield
   {title} {\bibinfo {title} {{Collider Probes of Axion-Like Particles}},\
  }\href {https://doi.org/10.1007/JHEP12(2017)044} {\bibfield  {journal}
  {\bibinfo  {journal} {JHEP}\ }\textbf {\bibinfo {volume} {12}},\ \bibinfo
  {pages} {044}},\ \Eprint {https://arxiv.org/abs/1708.00443} {arXiv:1708.00443
  [hep-ph]} \BibitemShut {NoStop}%
\bibitem [{\citenamefont {Heeck}\ and\ \citenamefont
  {Patel}(2019)}]{Heeck:2019guh}%
  \BibitemOpen
  \bibfield  {author} {\bibinfo {author} {\bibfnamefont {J.}~\bibnamefont
  {Heeck}}\ and\ \bibinfo {author} {\bibfnamefont {H.~H.}\ \bibnamefont
  {Patel}},\ }\bibfield  {title} {\bibinfo {title} {{Majoron at two loops}},\
  }\href {https://doi.org/10.1103/PhysRevD.100.095015} {\bibfield  {journal}
  {\bibinfo  {journal} {Phys. Rev. D}\ }\textbf {\bibinfo {volume} {100}},\
  \bibinfo {pages} {095015} (\bibinfo {year} {2019})},\ \Eprint
  {https://arxiv.org/abs/1909.02029} {arXiv:1909.02029 [hep-ph]} \BibitemShut
  {NoStop}%
\bibitem [{\citenamefont {Marsh}(2016)}]{Marsh:2015xka}%
  \BibitemOpen
  \bibfield  {author} {\bibinfo {author} {\bibfnamefont {D.~J.~E.}\
  \bibnamefont {Marsh}},\ }\bibfield  {title} {\bibinfo {title} {{Axion
  Cosmology}},\ }\href {https://doi.org/10.1016/j.physrep.2016.06.005}
  {\bibfield  {journal} {\bibinfo  {journal} {Phys. Rept.}\ }\textbf {\bibinfo
  {volume} {643}},\ \bibinfo {pages} {1} (\bibinfo {year} {2016})},\ \Eprint
  {https://arxiv.org/abs/1510.07633} {arXiv:1510.07633 [astro-ph.CO]}
  \BibitemShut {NoStop}%
\bibitem [{\citenamefont {Raffelt}\ and\ \citenamefont
  {Weiss}(1995)}]{Raffelt:1994ry}%
  \BibitemOpen
  \bibfield  {author} {\bibinfo {author} {\bibfnamefont {G.}~\bibnamefont
  {Raffelt}}\ and\ \bibinfo {author} {\bibfnamefont {A.}~\bibnamefont
  {Weiss}},\ }\bibfield  {title} {\bibinfo {title} {{Red giant bound on the
  axion - electron coupling revisited}},\ }\href
  {https://doi.org/10.1103/PhysRevD.51.1495} {\bibfield  {journal} {\bibinfo
  {journal} {Phys. Rev. D}\ }\textbf {\bibinfo {volume} {51}},\ \bibinfo
  {pages} {1495} (\bibinfo {year} {1995})},\ \Eprint
  {https://arxiv.org/abs/hep-ph/9410205} {arXiv:hep-ph/9410205} \BibitemShut
  {NoStop}%
\bibitem [{\citenamefont {Raffelt}(2008)}]{Raffelt:2006cw}%
  \BibitemOpen
  \bibfield  {author} {\bibinfo {author} {\bibfnamefont {G.~G.}\ \bibnamefont
  {Raffelt}},\ }\bibfield  {title} {\bibinfo {title} {{Astrophysical axion
  bounds}},\ }\href {https://doi.org/10.1007/978-3-540-73518-2_3} {\bibfield
  {journal} {\bibinfo  {journal} {Lect. Notes Phys.}\ }\textbf {\bibinfo
  {volume} {741}},\ \bibinfo {pages} {51} (\bibinfo {year} {2008})},\ \Eprint
  {https://arxiv.org/abs/hep-ph/0611350} {arXiv:hep-ph/0611350} \BibitemShut
  {NoStop}%
\bibitem [{\citenamefont {Caputo}\ and\ \citenamefont
  {Raffelt}(2024)}]{Caputo:2024oqc}%
  \BibitemOpen
  \bibfield  {author} {\bibinfo {author} {\bibfnamefont {A.}~\bibnamefont
  {Caputo}}\ and\ \bibinfo {author} {\bibfnamefont {G.}~\bibnamefont
  {Raffelt}},\ }\bibfield  {title} {\bibinfo {title} {{Astrophysical Axion
  Bounds: The 2024 Edition}},\ }\href {https://doi.org/10.22323/1.454.0041}
  {\bibfield  {journal} {\bibinfo  {journal} {PoS}\ }\textbf {\bibinfo {volume}
  {COSMICWISPers}},\ \bibinfo {pages} {041} (\bibinfo {year} {2024})},\ \Eprint
  {https://arxiv.org/abs/2401.13728} {arXiv:2401.13728 [hep-ph]} \BibitemShut
  {NoStop}%
\bibitem [{\citenamefont {Bollig}\ \emph {et~al.}(2020)\citenamefont {Bollig},
  \citenamefont {DeRocco}, \citenamefont {Graham},\ and\ \citenamefont
  {Janka}}]{Bollig:2020xdr}%
  \BibitemOpen
  \bibfield  {author} {\bibinfo {author} {\bibfnamefont {R.}~\bibnamefont
  {Bollig}}, \bibinfo {author} {\bibfnamefont {W.}~\bibnamefont {DeRocco}},
  \bibinfo {author} {\bibfnamefont {P.~W.}\ \bibnamefont {Graham}},\ and\
  \bibinfo {author} {\bibfnamefont {H.-T.}\ \bibnamefont {Janka}},\ }\bibfield
  {title} {\bibinfo {title} {{Muons in Supernovae: Implications for the
  Axion-Muon Coupling}},\ }\href
  {https://doi.org/10.1103/PhysRevLett.125.051104} {\bibfield  {journal}
  {\bibinfo  {journal} {Phys. Rev. Lett.}\ }\textbf {\bibinfo {volume} {125}},\
  \bibinfo {pages} {051104} (\bibinfo {year} {2020})},\ \bibinfo {note}
  {[Erratum: Phys.Rev.Lett. 126, 189901 (2021)]},\ \Eprint
  {https://arxiv.org/abs/2005.07141} {arXiv:2005.07141 [hep-ph]} \BibitemShut
  {NoStop}%
\bibitem [{\citenamefont {Croon}\ \emph {et~al.}(2021)\citenamefont {Croon},
  \citenamefont {Elor}, \citenamefont {Leane},\ and\ \citenamefont
  {McDermott}}]{Croon:2020lrf}%
  \BibitemOpen
  \bibfield  {author} {\bibinfo {author} {\bibfnamefont {D.}~\bibnamefont
  {Croon}}, \bibinfo {author} {\bibfnamefont {G.}~\bibnamefont {Elor}},
  \bibinfo {author} {\bibfnamefont {R.~K.}\ \bibnamefont {Leane}},\ and\
  \bibinfo {author} {\bibfnamefont {S.~D.}\ \bibnamefont {McDermott}},\
  }\bibfield  {title} {\bibinfo {title} {{Supernova Muons: New Constraints on
  $Z$' Bosons, Axions and ALPs}},\ }\href
  {https://doi.org/10.1007/JHEP01(2021)107} {\bibfield  {journal} {\bibinfo
  {journal} {JHEP}\ }\textbf {\bibinfo {volume} {01}},\ \bibinfo {pages}
  {107}},\ \Eprint {https://arxiv.org/abs/2006.13942} {arXiv:2006.13942
  [hep-ph]} \BibitemShut {NoStop}%
\bibitem [{\citenamefont {Navas}\ \emph {et~al.}(2024)\citenamefont {Navas}
  \emph {et~al.}}]{ParticleDataGroup:2024cfk}%
  \BibitemOpen
  \bibfield  {author} {\bibinfo {author} {\bibfnamefont {S.}~\bibnamefont
  {Navas}} \emph {et~al.} (\bibinfo {collaboration} {Particle Data Group}),\
  }\bibfield  {title} {\bibinfo {title} {{Review of particle physics}},\ }\href
  {https://doi.org/10.1103/PhysRevD.110.030001} {\bibfield  {journal} {\bibinfo
   {journal} {Phys. Rev. D}\ }\textbf {\bibinfo {volume} {110}},\ \bibinfo
  {pages} {030001} (\bibinfo {year} {2024})}\BibitemShut {NoStop}%
\bibitem [{\citenamefont {Jaffe}(1977)}]{Jaffe:1976yi}%
  \BibitemOpen
  \bibfield  {author} {\bibinfo {author} {\bibfnamefont {R.~L.}\ \bibnamefont
  {Jaffe}},\ }\bibfield  {title} {\bibinfo {title} {{Perhaps a Stable
  Dihyperon}},\ }\href {https://doi.org/10.1103/PhysRevLett.38.195} {\bibfield
  {journal} {\bibinfo  {journal} {Phys. Rev. Lett.}\ }\textbf {\bibinfo
  {volume} {38}},\ \bibinfo {pages} {195} (\bibinfo {year} {1977})},\ \bibinfo
  {note} {[Erratum: Phys.Rev.Lett. 38, 617 (1977)]}\BibitemShut {NoStop}%
\bibitem [{\citenamefont {Farrar}\ and\ \citenamefont
  {Zaharijas}(2004)}]{Farrar:2003qy}%
  \BibitemOpen
  \bibfield  {author} {\bibinfo {author} {\bibfnamefont {G.~R.}\ \bibnamefont
  {Farrar}}\ and\ \bibinfo {author} {\bibfnamefont {G.}~\bibnamefont
  {Zaharijas}},\ }\bibfield  {title} {\bibinfo {title} {{Nuclear and nucleon
  transitions of the H dibaryon}},\ }\href
  {https://doi.org/10.1103/PhysRevD.70.014008} {\bibfield  {journal} {\bibinfo
  {journal} {Phys. Rev. D}\ }\textbf {\bibinfo {volume} {70}},\ \bibinfo
  {pages} {014008} (\bibinfo {year} {2004})},\ \Eprint
  {https://arxiv.org/abs/hep-ph/0308137} {arXiv:hep-ph/0308137} \BibitemShut
  {NoStop}%
\bibitem [{\citenamefont {Farrar}(2018)}]{Farrar:2017ysn}%
  \BibitemOpen
  \bibfield  {author} {\bibinfo {author} {\bibfnamefont {G.~R.}\ \bibnamefont
  {Farrar}},\ }\bibfield  {title} {\bibinfo {title} {{6-quark Dark Matter}},\
  }\href {https://doi.org/10.22323/1.301.0929} {\bibfield  {journal} {\bibinfo
  {journal} {PoS}\ }\textbf {\bibinfo {volume} {ICRC2017}},\ \bibinfo {pages}
  {929} (\bibinfo {year} {2018})},\ \Eprint {https://arxiv.org/abs/1711.10971}
  {arXiv:1711.10971 [hep-ph]} \BibitemShut {NoStop}%
\bibitem [{\citenamefont {Kolb}\ and\ \citenamefont
  {Turner}(2019)}]{Kolb:2018bxv}%
  \BibitemOpen
  \bibfield  {author} {\bibinfo {author} {\bibfnamefont {E.~W.}\ \bibnamefont
  {Kolb}}\ and\ \bibinfo {author} {\bibfnamefont {M.~S.}\ \bibnamefont
  {Turner}},\ }\bibfield  {title} {\bibinfo {title} {{Dibaryons cannot be the
  dark matter}},\ }\href {https://doi.org/10.1103/PhysRevD.99.063519}
  {\bibfield  {journal} {\bibinfo  {journal} {Phys. Rev. D}\ }\textbf {\bibinfo
  {volume} {99}},\ \bibinfo {pages} {063519} (\bibinfo {year} {2019})},\
  \Eprint {https://arxiv.org/abs/1809.06003} {arXiv:1809.06003 [hep-ph]}
  \BibitemShut {NoStop}%
\bibitem [{\citenamefont {McDermott}\ \emph {et~al.}(2019)\citenamefont
  {McDermott}, \citenamefont {Reddy},\ and\ \citenamefont
  {Sen}}]{McDermott:2018ofd}%
  \BibitemOpen
  \bibfield  {author} {\bibinfo {author} {\bibfnamefont {S.~D.}\ \bibnamefont
  {McDermott}}, \bibinfo {author} {\bibfnamefont {S.}~\bibnamefont {Reddy}},\
  and\ \bibinfo {author} {\bibfnamefont {S.}~\bibnamefont {Sen}},\ }\bibfield
  {title} {\bibinfo {title} {{Deeply bound dibaryon is incompatible with
  neutron stars and supernovae}},\ }\href
  {https://doi.org/10.1103/PhysRevD.99.035013} {\bibfield  {journal} {\bibinfo
  {journal} {Phys. Rev. D}\ }\textbf {\bibinfo {volume} {99}},\ \bibinfo
  {pages} {035013} (\bibinfo {year} {2019})},\ \Eprint
  {https://arxiv.org/abs/1809.06765} {arXiv:1809.06765 [hep-ph]} \BibitemShut
  {NoStop}%
\end{thebibliography}%

\end{document}